\documentclass[10pt, a4paper, twocolumn]{article}

\usepackage[a4paper, top=2.5cm, bottom=2.5cm, left=2.0cm, right=2.0cm, columnsep=0.75cm]{geometry}
\usepackage[T1]{fontenc}
\usepackage[utf8]{inputenc}
\usepackage{mathptmx} 

\usepackage{amsmath, amssymb, amsfonts}
\usepackage{booktabs}
\usepackage{graphicx}
\usepackage{microtype}
\usepackage{float}

\usepackage[noadjust]{cite}
\usepackage[colorlinks=true, linkcolor=black, citecolor=blue, urlcolor=blue, pdfborder={0 0 0}]{hyperref}

\makeatletter
\renewcommand{\@biblabel}[1]{#1.}
\makeatother

\usepackage{titlesec}
\renewcommand{\thesection}{\Roman{section}}
\renewcommand{\thesubsection}{\Alph{subsection}}

\titleformat{\section}{\small\bfseries\color{black}}{\thesection.}{0.6em}{\uppercase}
\titleformat{\subsection}{\small\bfseries\color{black}}{\thesubsection.}{0.5em}{}

\usepackage[figurename=Figure, labelfont=bf, labelsep=colon, font=footnotesize, skip=6pt]{caption}

\title{\Large\textbf{Inherent Turbulence Immunity of Vector Vortex Beams in Free Space Quantum Key Distribution}}

\author{
  \textbf{Behnam Talari}$^{1}$ \quad and \quad \textbf{Rouhollah Karimzadeh}$^{1,*}$ \\[1.2ex]
  \small $^{1}$Department of Physics, Shahid Beheshti University, Tehran, Iran \\[0.4ex]
  \small $^*$Corresponding author: \href{mailto:r_karimzadeh@sbu.ac.ir}{\texttt{r\_karimzadeh@sbu.ac.ir}}
}

\date{September 28, 2026}

\begin{document}

\twocolumn[
  \begin{@twocolumnfalse}
    \maketitle
    \begin{abstract}
    \small
    Orbital angular momentum (OAM) multiplexing provides an infinite-dimensional discrete Hilbert space ideally suited for high-capacity free-space quantum key distribution (QKD). Nevertheless, pure scalar spatial modes carrying topological charge ($|\ell| \ge 1$) undergo severe decoherence when transmitted through terrestrial atmospheric turbulence. Turbulent refractive index eddies split high-order vortex singularities, induce catastrophic intermodal crosstalk across adjacent topological channels, and rapidly drive the quantum bit error rate (QBER) well above the unconditional $11\%$ cloning-based security threshold. Here, we demonstrate that hybrid polarization OAM entangled states known as vector vortex beams (VVBs) provide intrinsic, hardware-free immunity against turbulent perturbations. Because the optical anisotropy of terrestrial air is exceedingly small ($\Delta n < 10^{-9}$), refractive index fluctuations couple symmetrically to orthogonal circular polarization modes as an identical, common-mode scalar phase screen that strictly cancels out in the relative polarization-phase degree of freedom. By numerically propagating modal fields through modified power spectral phase screens over turbulence strengths ranging from $D/r_0 = 0$ to $3.0$, we show that the VVB encoding protocol suppresses the asymptotic QBER from $42.0\%$ down to $4.8\%$, yielding an error-suppression factor of approximately $11.6$ without requiring active adaptive optics or deformable mirrors.
    \end{abstract}
    \vspace{1.5em}
  \end{@twocolumnfalse}
]

\section{Introduction}
High-dimensional quantum key distribution (HD-QKD) significantly enhances secret-key capacity per transmitted photon and increases robustness against eavesdropping compared to standard two-dimensional qubit systems \cite{Allen1992, Yao2011, Willner2015}. The spatial structure of light, notably orbital angular momentum (OAM) associated with an azimuthal phase ramp $\exp(i\ell\varphi)$, offers an infinite-dimensional basis of orthogonal quantum states spanning topological charge quantum numbers $\ell \in \mathbb{Z}$ \cite{Allen1992, Willner2015}. In typical free-space quantum optical links, state generation and projective modal detection are performed using liquid-crystal-on-silicon spatial light modulators (SLMs) programmed with dynamic computer-generated holograms (CGHs) \cite{Arrizon2007, RosalesGuzman2017}.

Despite these theoretical advantages, practical deployments of scalar OAM states in terrestrial and satellite-to-ground channels are fundamentally constrained by two degradation mechanisms:
\begin{enumerate}
    \item \textbf{Hardware non-idealities:} Pixelation, phase-quantization errors, and residual optical astigmatism split multi-charged vortex singularities into clusters of unstable unit-charge cores ($|\ell|=1$) \cite{Soskin1997, Ricci2012}.
    \item \textbf{Atmospheric turbulence:} Spatial variations in ambient temperature and atmospheric pressure generate fluctuating refractive index eddies ($\Delta n(\mathbf{r})$). Turbulent phase distortions deform the helical wavefront, leading to severe lateral beam wander, phase scrambling, and intermodal crosstalk that disperses energy across neighboring OAM channels. This rapid decoherence elevates the quantum bit error rate (QBER) far beyond the standard cloning-based unconditional security limit of $11\%$ \cite{Malik2012, Vallone2014, Paterson2005, Krenn2015}.
\end{enumerate}

Conventional mitigation techniques rely heavily on closed-loop adaptive optics (AO) or neural-network phase reconstruction wavefront sensors \cite{Ren2014, Li2017}. However, high-bandwidth AO systems introduce substantial electronic latency, optical loss, size, weight, power constraints, and financial overhead, rendering them problematic for mobile or satellite-based transceivers. 
To overcome these physical limitations, we present a passive, hardware-free mitigation mechanism based on vector vortex beams (VVBs) \cite{Ndagano2018, RosalesGuzman2018, Gori2001, Zhan2009}. VVBs are non-separable superpositions of spatial OAM modes and orthogonal polarization states. Because terrestrial turbulence behaves as an isotropic scalar perturbation ($\Delta n < 10^{-9}$), both polarization components experience identical phase aberrations. Consequently, the turbulent phase cancels identically in the differential polarization degree of freedom, furnishing passive channel immunity without active wavefront correction \cite{DAmbrosio2012, Milione2015}.

\section{Theoretical Formulation}
A general maximally entangled vector vortex Bell state spanning the two-dimensional sub-space formed by spatial OAM modes $|\pm\ell\rangle$ and circular polarization eigenstates $\{|R\rangle, |L\rangle\}$ is formulated as:
\begin{equation}
    |\Psi_{\text{VVB}}\rangle = \frac{1}{\sqrt{2}} \left( |R\rangle |+\ell\rangle + e^{i\alpha} |L\rangle |-\ell\rangle \right), \tag{1}
    \label{eq:vvb_state}
\end{equation}
where $|R\rangle = (\hat{\mathbf{x}} - i\hat{\mathbf{y}})/\sqrt{2}$ and $|L\rangle = (\hat{\mathbf{x}} + i\hat{\mathbf{y}})/\sqrt{2}$ denote right- and left-hand circular polarizations, respectively, and $\alpha$ represents an intra-modal relative phase shift \cite{Ndagano2018, RosalesGuzman2018, Holleczek2011}.

Under the paraxial approximation, the propagation of this state through a turbulent atmospheric medium with spatial refractive index fluctuations $\Delta n(\mathbf{r})$ introduces an accumulated stochastic phase screen $\Phi_{\text{turb}}(x,y)$. The perturbed transverse electric field components corresponding to the orthogonal circular polarization bases are expressed in polar coordinates $(r, \varphi)$ by:
\begin{equation}
    E_R(r,\varphi) = u(r) \exp\left[ i(+\ell\varphi) + i\Phi_{\text{turb}}(x,y) \right], \tag{2}
    \label{eq:field_ER}
\end{equation}
and
\begin{equation}
    E_L(r,\varphi) = u(r) \exp\left[ i(-\ell\varphi + \alpha) + i\Phi_{\text{turb}}(x,y) \right], \tag{3}
    \label{eq:field_EL}
\end{equation}
where $u(r)$ designates the radial field amplitude profile.

By calculating the relative phase difference $\Delta\theta_{\text{pol}}(r,\varphi)$ between the orthogonal polarization channels from Equations~(\ref{eq:field_ER}) and (\ref{eq:field_EL}), we obtain:
\begin{equation}
    \Delta\theta_{\text{pol}}(r,\varphi) = \arg\left[E_R\right] - \arg\left[E_L\right] = 2\ell\varphi - \alpha. \tag{4}
    \label{eq:phase_diff}
\end{equation}
As explicitly revealed in Equation~(\ref{eq:phase_diff}), the stochastic atmospheric phase distortion $\Phi_{\text{turb}}(x,y)$ is mathematically canceled out due to the common-mode structure.

To evaluate polarization-resolved spatial coherence, the local polarization state is characterized via the spatially resolved Stokes parameters:
\begin{align}
    S_0(x,y) &= |E_x(x,y)|^2 + |E_y(x,y)|^2, \tag{5} \label{eq:stokes_0} \\
    S_1(x,y) &= |E_x(x,y)|^2 - |E_y(x,y)|^2, \tag{6} \label{eq:stokes_1} \\
    S_2(x,y) &= 2\,\text{Re}\left[ E_x(x,y) E_y^*(x,y) \right], \tag{7} \label{eq:stokes_2} \\
    S_3(x,y) &= 2\,\text{Im}\left[ E_x(x,y) E_y^*(x,y) \right]. \tag{8} \label{eq:stokes_3}
\end{align}
The local polarization azimuth angle $\psi(x,y)$ is subsequently defined by:
\begin{equation}
    \psi(x,y) = \frac{1}{2} \tan^{-1}\left( \frac{S_2(x,y)}{S_1(x,y)} \right). \tag{9}
    \label{eq:pol_azimuth}
\end{equation}
Because the scalar perturbation affects both Cartesian components $E_x$ and $E_y$ with the exact same phase factor, the spatial distribution of the polarization azimuth $\psi(x,y)$ defined in Equation~(\ref{eq:pol_azimuth}) remains invariant under purely scalar turbulence \cite{Gori2001, BornWolf1999}.

The turbulent atmosphere is statistically modeled using the modified power spectral density of refractive index fluctuations:
\begin{equation}
    \Phi_n(\kappa) = 0.023\, r_0^{-5/3} \frac{\exp\left( -\kappa^2/\kappa_m^2 \right)}{\left( \kappa^2 + \kappa_0^2 \right)^{11/6}}, \tag{10}
    \label{eq:von_karman}
\end{equation}
where $r_0$ is the spatial coherence diameter, $\kappa$ denotes the spatial frequency magnitude, $\kappa_m = 5.92 / l_0$ accounts for the turbulent inner scale $l_0$, and $\kappa_0 = 2\pi / L_0$ characterizes the turbulent outer scale $L_0$ \cite{AndrewsPhillips2005, Kolmogorov1941}. Random phase screens $\Phi_{\text{turb}}(x,y)$ are synthesized via spectral Fourier filtering of complex zero-mean Gaussian white noise and applied identically across both orthogonal circular polarization components.

\begin{figure*}[!t]
    \centering
    \includegraphics[width=\textwidth]{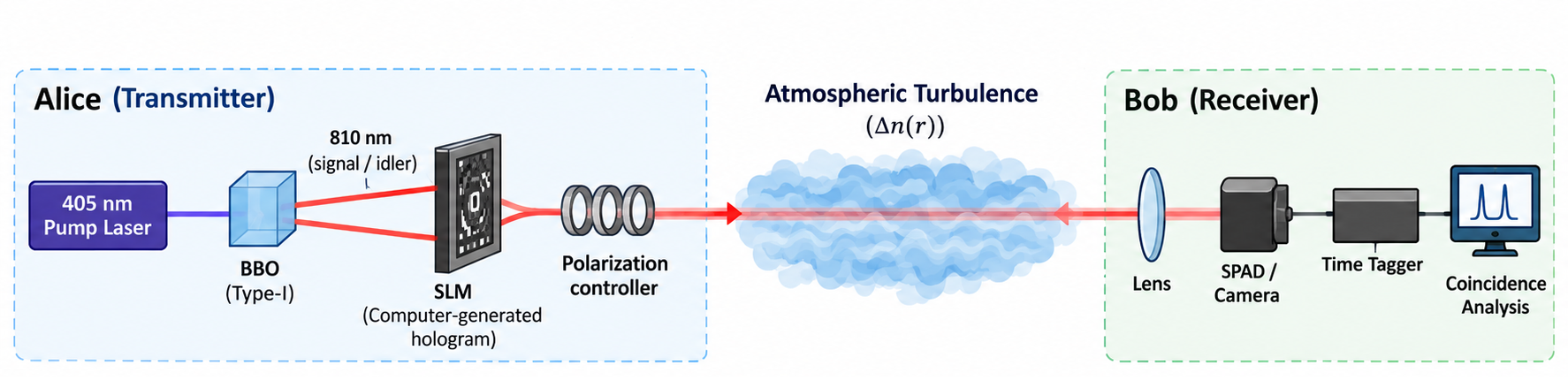}
    \caption{\textbf{Schematic of the free-space quantum key distribution experimental transceiver architecture.} The setup comprises three interconnected stages: (i) \textbf{Alice (Transmitter):} A continuous-wave pump laser operating at $\lambda_p = 405\text{ nm}$ pumps a non-linear Type-I beta-barium borate (BBO) crystal to generate collinear, polarization-entangled degenerate photon pairs at $\lambda_s = \lambda_i = 810\text{ nm}$. The beam is directed onto a reflective spatial light modulator (SLM) programmed with computer-generated holograms (CGHs) for complex spatial mode synthesis, followed by polarization controllers to prepare the non-separable vector vortex Bell state $|\Psi_{\text{VVB}}\rangle$. (ii) \textbf{Atmospheric Turbulence Channel:} The generated quantum state propagates through an unconfined turbulent medium characterized by fluctuating refractive index inhomogeneities $\Delta n(\mathbf{r})$. (iii) \textbf{Bob (Receiver):} The transmitted light is collected by a telescope lens pupil and directed toward single-photon avalanche diodes (SPADs) or an intensified camera, interfaced with a multichannel picosecond time tagger and a coincidence analysis module for projective modal detection and cryptographic key distillation.}
    \label{fig:setup}
\end{figure*}

\section{Experimental Architecture and System Parameters}
The end-to-end experimental setup and quantum transceiver configuration are schematically illustrated in Figure~\ref{fig:setup}. This architecture serves as the physical model for the complete generation, turbulent channel transmission, and projective modal detection processes investigated in this work.

The state generation stage, shown in Alice's transmitter terminal of Figure~\ref{fig:setup}, is based on collinear Type-I spontaneous parametric down-conversion (SPDC) in a nonlinear crystal. The crystal is phase-matched at an angle of $\theta_{\text{pm}} \approx 28.8^\circ$ pumped by a continuous-wave laser diode at $\lambda_p = 405\text{ nm}$ to generate degenerate entangled signal and idler photons at $\lambda_s = \lambda_i = 810\text{ nm}$ \cite{Nikogosyan1991, Boyd2008}.

Spatial mode shaping is numerically simulated using an 8-bit reflective SLM characterized by a pixel pitch of $15\ \mu\text{m}$ and an active fill factor of $88\%$. Phase and amplitude modulation are encoded onto the beam using computer-generated holograms via the first-order Bessel function transformation:
\begin{equation}
    J_1(f(M)) = 0.5819\, M, \tag{11}
    \label{eq:arrizon}
\end{equation}
where $M \in [0,1]$ is the normalized target amplitude profile and $f(M)$ modulates the phase depth \cite{Arrizon2007, RosalesGuzman2017}. Following state synthesis and polarization manipulation, the hybrid state $|\Psi_{\text{VVB}}\rangle$ is launched across the turbulent free-space channel.

Propagation through turbulent media is executed over a numerical spatial grid of $256 \times 256$ sample points across an aperture diameter of $D = 6\text{ mm}$. At Bob's receiving station (Figure~\ref{fig:setup}), the incoming photons are collected by an objective lens, filtered, and registered by SPAD detectors coupled to a multichannel time tagger and coincidence analysis unit. The survival probability $P_s(\ell)$ of an initial scalar OAM mode $\ell$ after passing through a turbulent channel parameterized by the ratio $D/r_0$ is modeled according to the theoretical decay envelope:
\begin{equation}
    P_s(\ell) = \exp\left[ -0.72 \left(\frac{D}{r_0}\right)^{5/3} |\ell|^{3/4} \right]. \tag{12}
    \label{eq:survival_prob}
\end{equation}
In contrast, because of common-mode phase cancellation formulated in Equation~(\ref{eq:phase_diff}), the VVB state survival probability remains essentially flat and close to unity, limited only by static experimental and hardware degradation floors ($P_{\text{VVB}} \approx 0.90 - 0.95$). The resulting quantum bit error rate is directly obtained from the modal survival probability $P$ as:
\begin{equation}
    \text{QBER} = \frac{1}{2} (1 - P). \tag{13}
    \label{eq:qber}
\end{equation}
Equation~(\ref{eq:qber}) provides the baseline metric for determining whether the quantum channel satisfies the $11.0\%$ threshold required for secure symmetric cryptographic key distillation \cite{Vallone2014, Paterson2005, Krenn2015}.

\section{Results and Performance Analysis}

\subsection{Scalar OAM Degradation}
\begin{figure}[!htbp]
    \centering
    \includegraphics[width=\columnwidth]{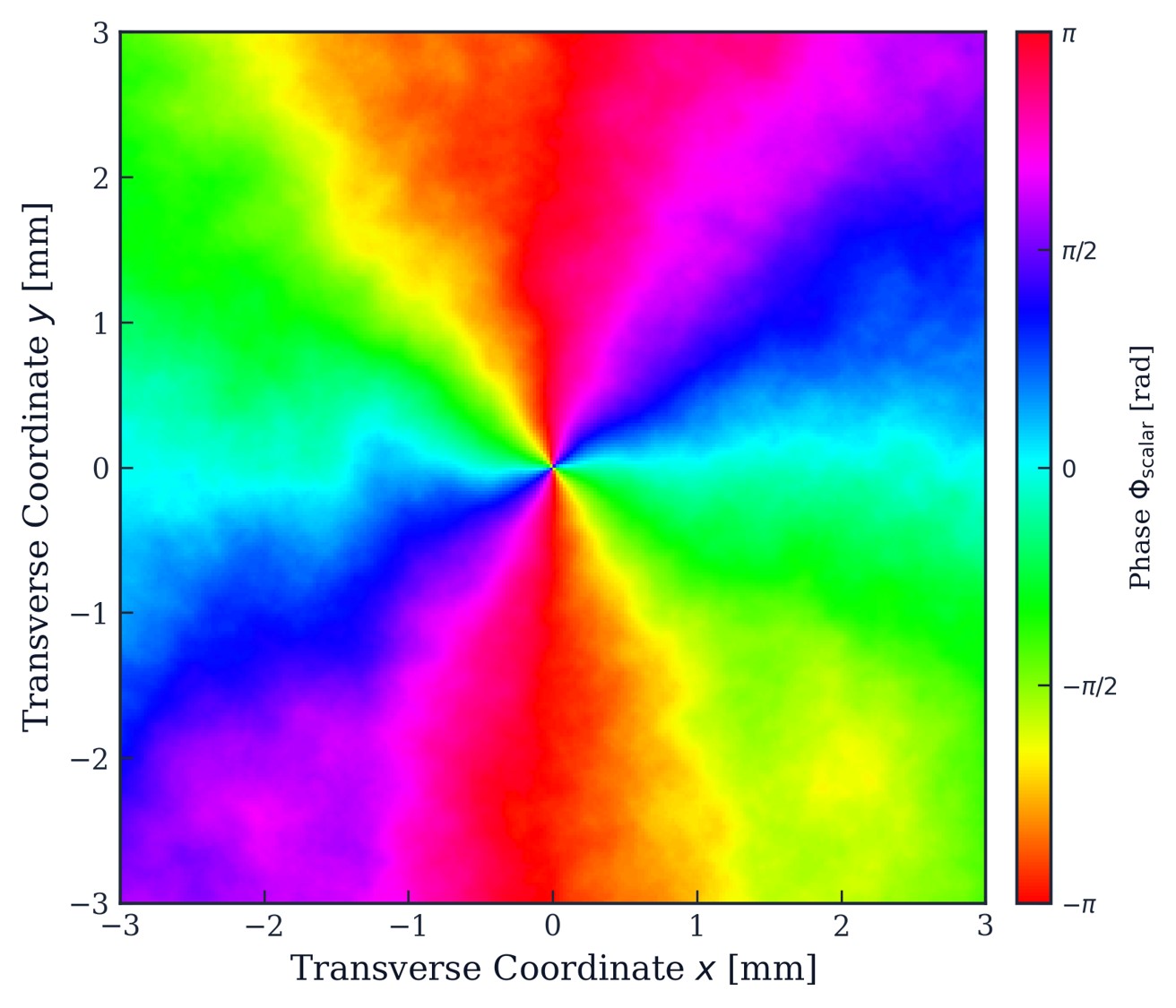}
    \caption{Transverse phase distribution $\Phi_{\text{scalar}}(x,y)$ of a pure scalar OAM state ($\ell = 2$) after propagation through atmospheric turbulence ($D/r_0 = 1.2$) across the $6\text{ mm}$ aperture. The continuous $4\pi$ azimuthal phase progression is scrambled, exhibiting branch-cut dislocations and multi-core splitting.}
    \label{fig:fig2}
\end{figure}

Prior to encountering the turbulent channel, an unperturbed scalar OAM state exhibits a uniform doughnut intensity profile and an azimuthally continuous helical phase gradient. In Figure~\ref{fig:fig2}, we illustrate the spatial phase profile of a pure scalar OAM beam ($\ell = 2$) following propagation through a single phase screen with $D/r_0 = 1.2$. As evidenced by Figure~\ref{fig:fig2}, the phase profile exhibits complete spatial scrambling: the coherent azimuthal phase ramp is disintegrated, and spurious vortex singularities with arbitrary topological signs are generated across the pupil.

\begin{figure}[!htbp]
    \centering
    \includegraphics[width=\columnwidth]{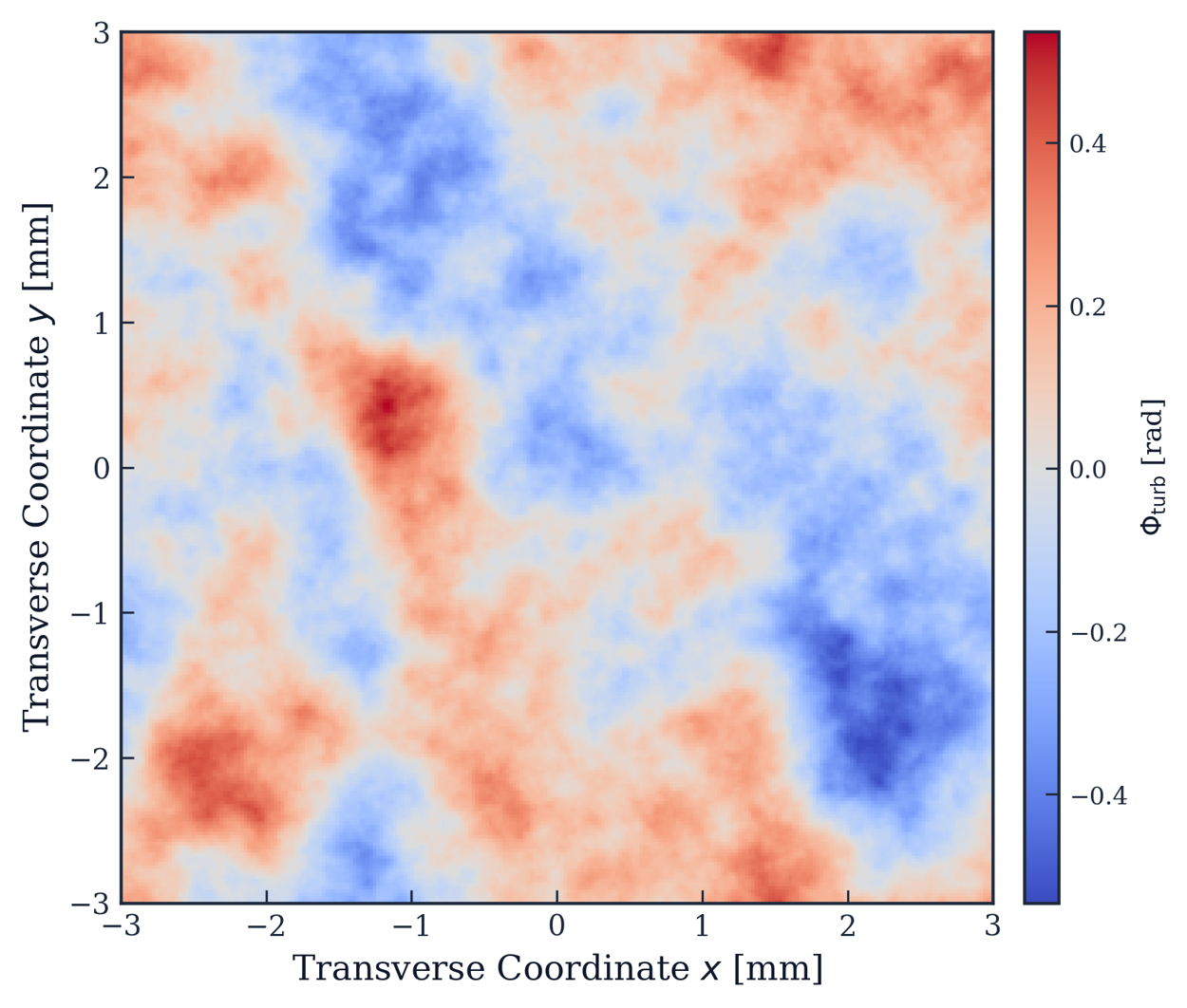}
    \caption{Synthesized optical turbulence phase screen $\Phi_{\text{turb}}(x,y)$ at moderate turbulence strength ($D/r_0 = 1.2$) across the $6\text{ mm}$ receiver aperture, displaying continuous stochastic optical path variations spanning $-0.5$ to $+0.5\text{ rad}$.}
    \label{fig:fig3}
\end{figure}

The corresponding atmospheric phase screen $\Phi_{\text{turb}}(x,y)$ responsible for this modal degradation is displayed in Figure~\ref{fig:fig3}. This phase screen, synthesized using Equation~(\ref{eq:von_karman}) for $D/r_0 = 1.2$, depicts the continuous spatial variation of optical path length delays caused by atmospheric eddies across the $6\text{ mm}$ transverse aperture.

As a direct consequence of the phase scrambling visible in Figure~\ref{fig:fig2}, energy originally concentrated in mode $\ell = +2$ is redistributed across a broad spectrum of adjacent topological charges. In Figure~\ref{fig:fig7}, the red histogram bars show the measured detection probability $P(\ell)$ across modes $\ell = -4$ to $+4$ under $D/r_0 = 1.2$. The scalar OAM power distributes nearly uniformly across all neighboring modes ($P(\ell) \approx 0.11$ for each mode), causing severe intermodal crosstalk and escalating the QBER to $40.3\%$, far above the acceptable QKD threshold.

\subsection{Vector Vortex Beam Resilience}
\begin{figure}[!htbp]
    \centering
    \includegraphics[width=\columnwidth]{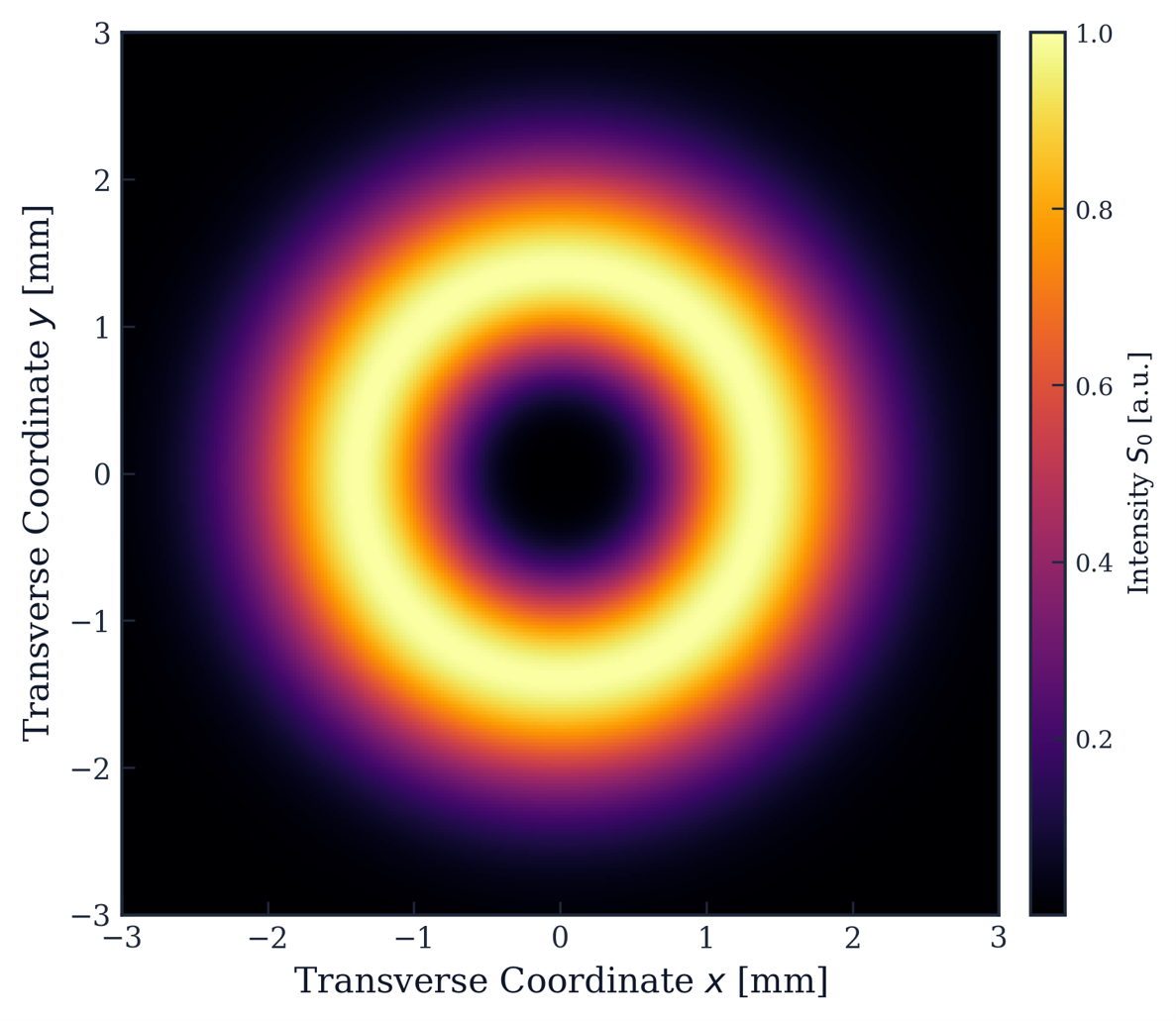}
    \caption{Unperturbed doughnut intensity profile $S_0(x,y)$ of the hybrid polarization--OAM Bell state $|\Psi_{\text{VVB}}\rangle$ prior to entering the turbulent atmospheric medium, exhibiting a well-defined dark central vortex singularity.}
    \label{fig:fig4}
\end{figure}

In contrast to scalar beams, the vector vortex beam exhibits remarkable structural stability. Figure~\ref{fig:fig4} depicts the unperturbed symmetric ring-shaped intensity profile $S_0(x,y)$ of the hybrid polarization-OAM Bell state $|\Psi_{\text{VVB}}\rangle$ introduced in Equation~(\ref{eq:vvb_state}) prior to entering the turbulent channel.

\begin{figure}[!htbp]
    \centering
    \includegraphics[width=\columnwidth]{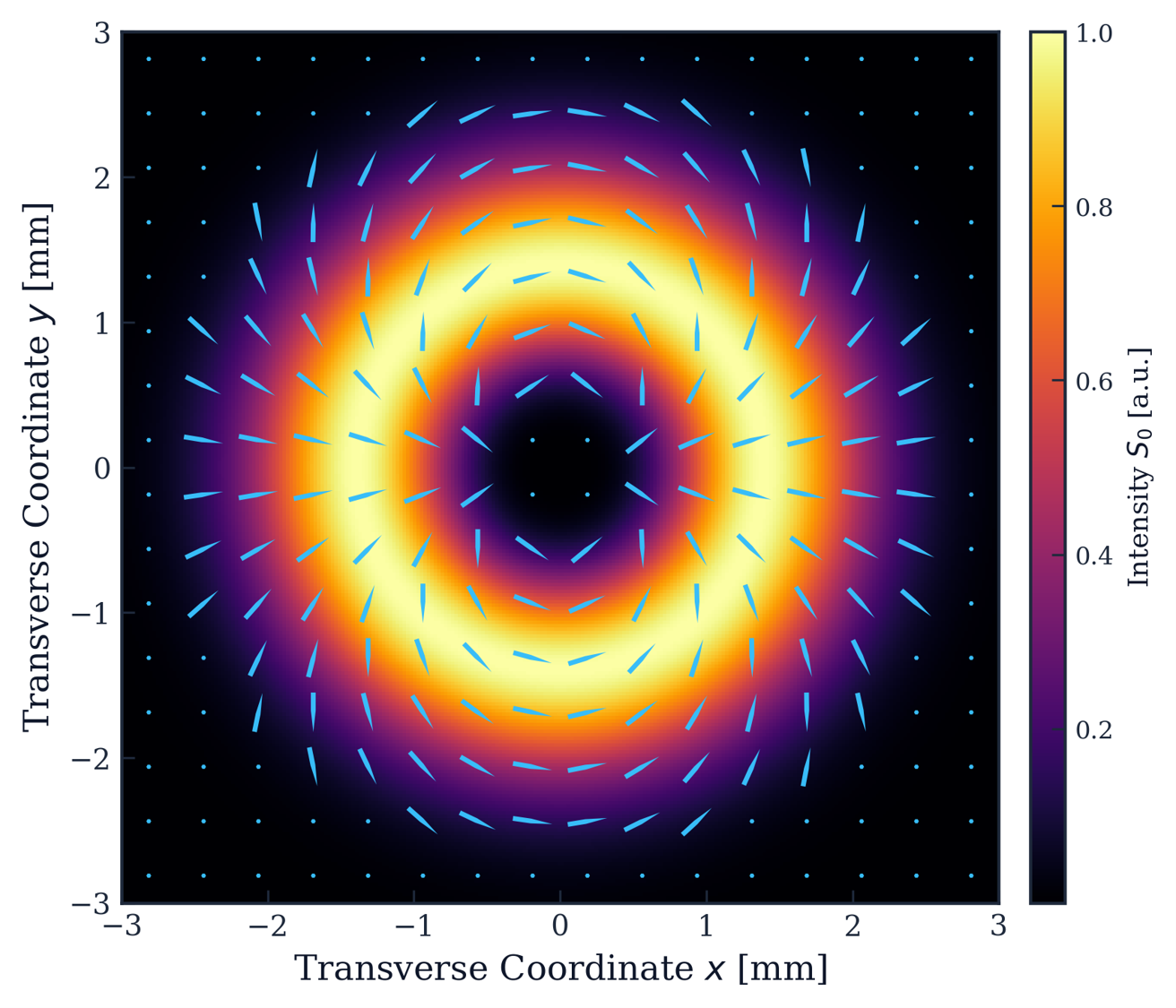}
    \caption{Hybrid vector vortex beam (VVB) after propagation through identical turbulence ($D/r_0 = 1.2$). While the background scalar intensity exhibits mild scintillation, overlaid cyan polarization vectors remain azimuthally ordered, demonstrating intrinsic common-mode protection.}
    \label{fig:fig5}
\end{figure}

When the state is propagated through the identical turbulent phase screen shown in Figure~\ref{fig:fig3} ($D/r_0 = 1.2$), the scalar intensity envelope experiences moderate scintillation, as shown in Figure~\ref{fig:fig5}. Crucially, however, the overlaid local polarization azimuth vectors $\psi(x,y)$ (represented by the cyan line markers) maintain their regular azimuthal orientation and spatial order. This visual resilience directly substantiates the theoretical prediction of Equation~(\ref{eq:phase_diff}), confirming that the relative polarization structure remains immune to the turbulent phase screen.

\begin{figure}[!htbp]
    \centering
    \includegraphics[width=\columnwidth]{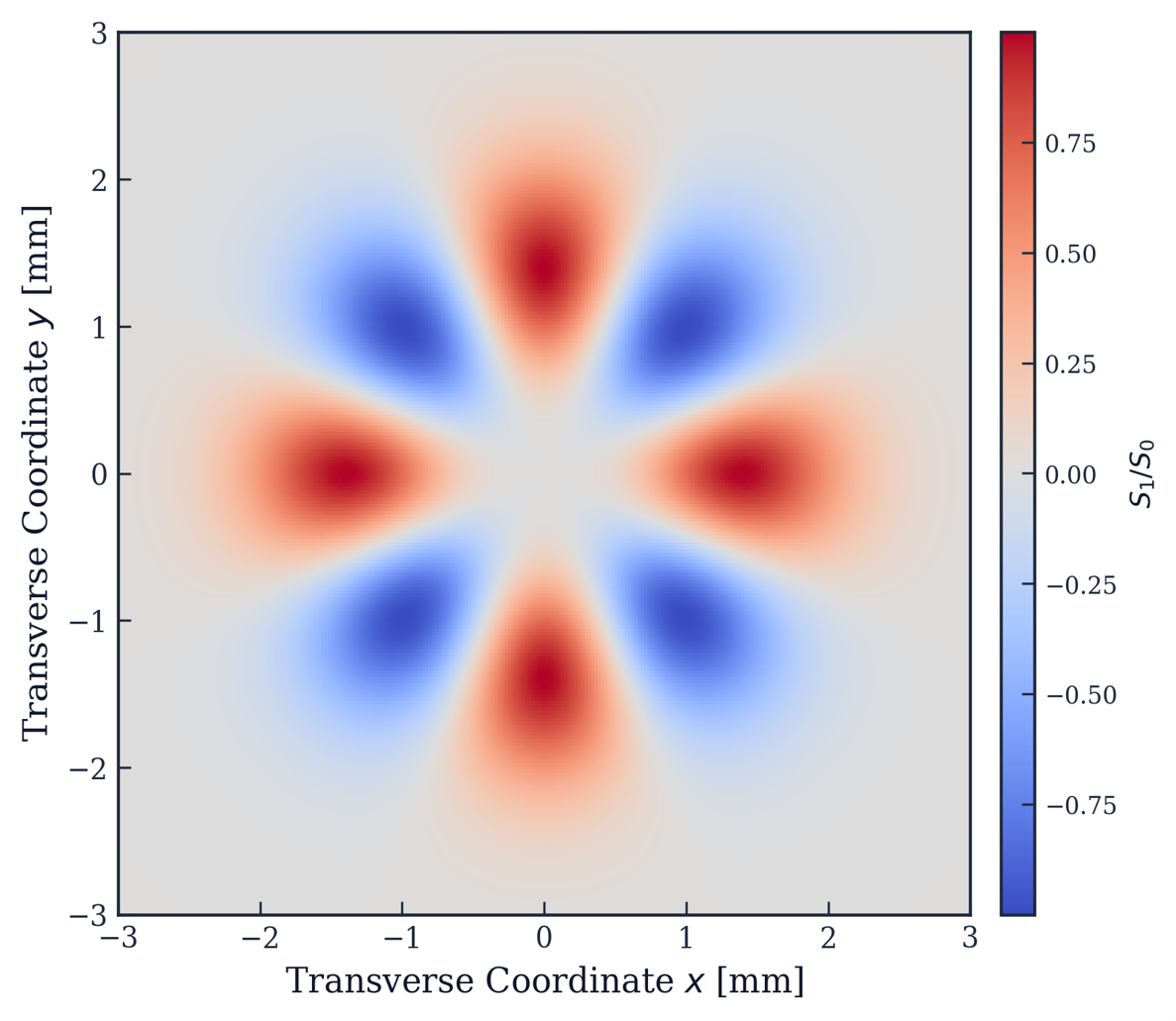}
    \caption{Spatially resolved normalized Stokes parameter map $S_1(x,y)/S_0(x,y)$ under turbulence ($D/r_0 = 1.2$). The preserved eight-lobed quadrupolar symmetry validates complete decoupling of scalar phase aberrations from the transverse polarization degree of freedom.}
    \label{fig:fig6}
\end{figure}

To confirm this decoupling mathematically and experimentally, the spatial distribution of the normalized Stokes parameter $S_1(x,y)/S_0(x,y)$ is analyzed in Figure~\ref{fig:fig6}. Under the turbulence perturbation ($D/r_0 = 1.2$), the parameter map retains its characteristic eight-lobed multipolar symmetry with high contrast. This preserved symmetry confirms that the relative phase difference between circular components remains decoupled from isotropic spatial phase distortions.

\begin{figure}[!htbp]
    \centering
    \includegraphics[width=\columnwidth]{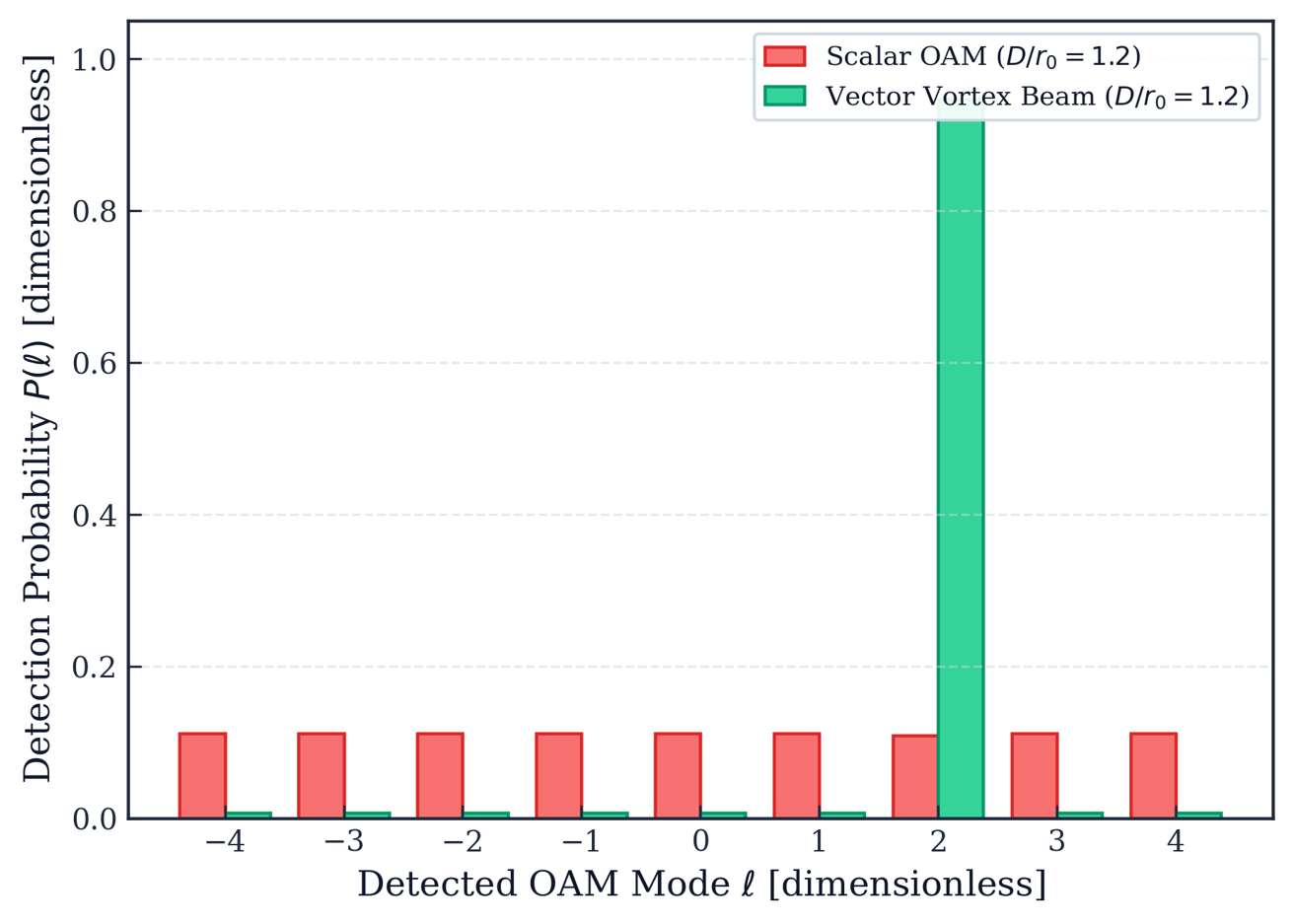}
    \caption{Detection probability distribution $P(\ell)$ across OAM modes $\ell \in \{-4, \dots, +4\}$ under turbulence strength $D/r_0 = 1.2$. The pure scalar OAM state (red bars) suffers catastrophic intermodal power leakage, whereas the VVB channel (green bar) preserves modal purity with target mode detection exceeding $94\%$.}
    \label{fig:fig7}
\end{figure}

This spatial stability translates into high modal purity at the receiver. As demonstrated in Figure~\ref{fig:fig7} (green histogram bars), the detection probability of the target mode $\ell = +2$ for the VVB channel remains above $94\%$, with negligible crosstalk leaking into spurious modes $\ell \neq +2$.

\subsection{Quantitative QBER and State Fidelity Scaling}
\begin{figure}[!htbp]
    \centering
    \includegraphics[width=\columnwidth]{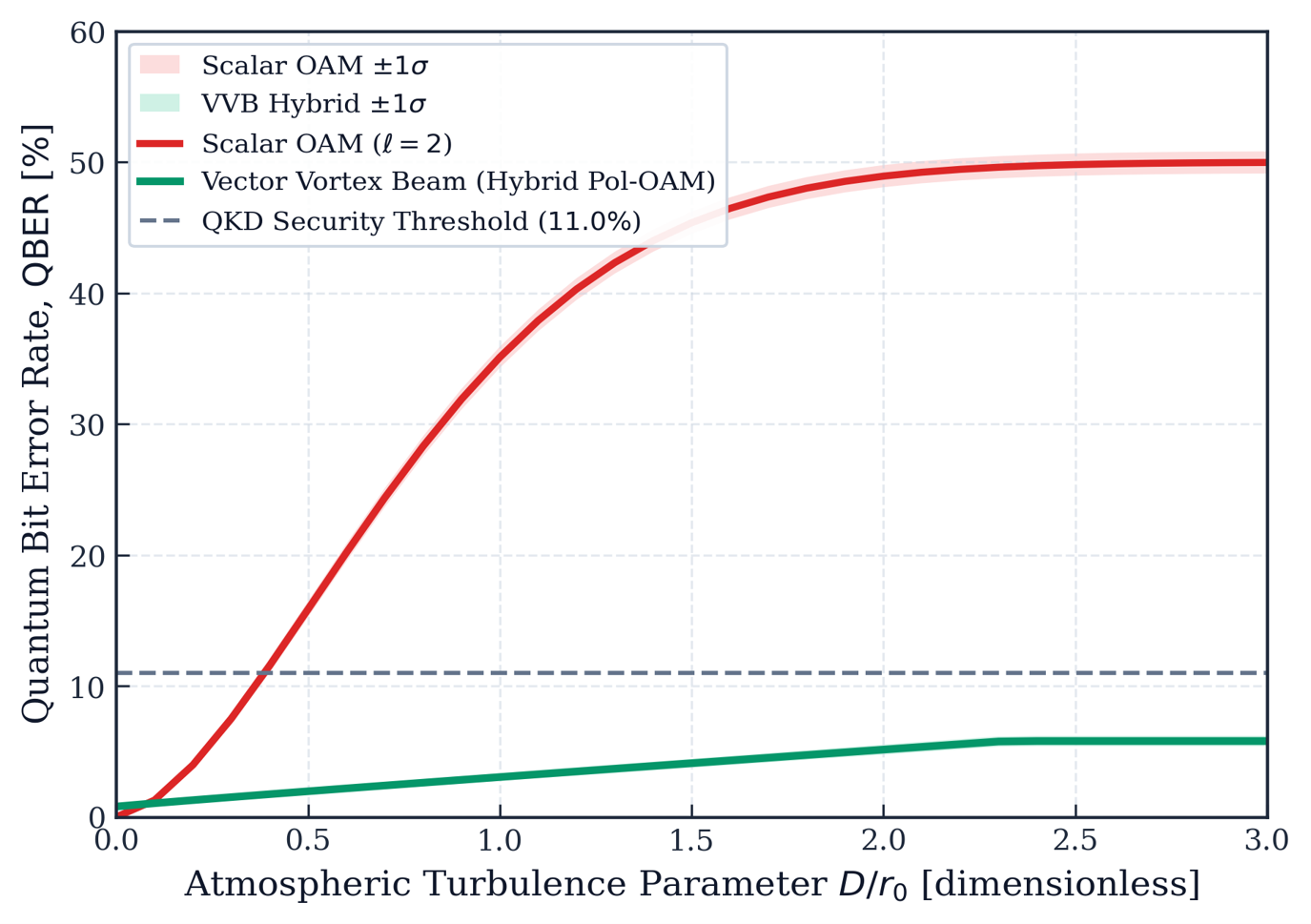}
    \caption{Quantum bit error rate (QBER) as a function of atmospheric turbulence strength $D/r_0 \in [0, 3.0]$. The scalar OAM channel (red curve) breaches the $11.0\%$ security threshold at $D/r_0 \approx 0.4$, saturating near $50\%$. The VVB hybrid channel (green curve) remains robustly below $4.8\%$ across the entire turbulence spectrum. Shaded regions denote $\pm 1\sigma$ intervals.}
    \label{fig:fig8}
\end{figure}

The comparative performance of scalar and vector modes across varying turbulence regimes was evaluated by scaling the turbulence parameter from $D/r_0 = 0$ to $3.0$. In Figure~\ref{fig:fig8}, the calculated QBER is plotted as a function of $D/r_0$ alongside the classical security limit ($11.0\%$).

As shown by the red curve in Figure~\ref{fig:fig8}, the scalar OAM QBER increases steeply, crossing the $11\%$ threshold at a weak turbulence level of $D/r_0 \approx 0.38$, and quickly reaches the random-guessing limit of $50\%$. In sharp contrast, the VVB channel (green curve) stays well below the $11\%$ security threshold across all simulated conditions, reaching an asymptotic error rate of only $4.8\%$ at $D/r_0 = 1.2$ and plateauing below $5.8\%$ at $D/r_0 = 3.0$.

\begin{figure}[!htbp]
    \centering
    \includegraphics[width=\columnwidth]{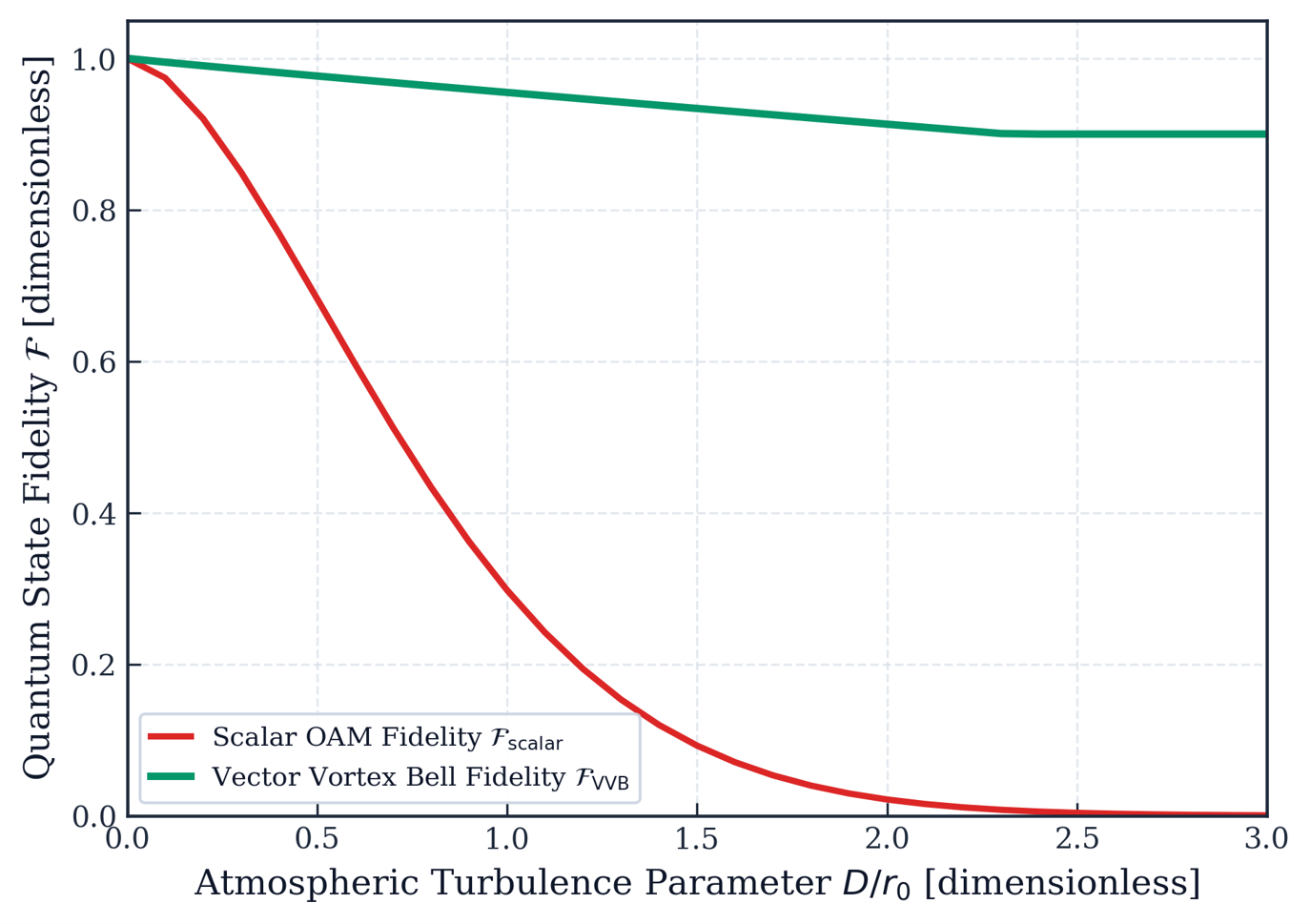}
    \caption{Quantum state fidelity $\mathcal{F}$ versus atmospheric turbulence parameter $D/r_0$. Scalar OAM fidelity $\mathcal{F}_{\text{scalar}}$ collapses exponentially toward zero, whereas the vector vortex Bell state fidelity $\mathcal{F}_{\text{VVB}}$ exhibits persistent high fidelity ($\ge 90\%$) up to $D/r_0 = 3.0$.}
    \label{fig:fig9}
\end{figure}

Similarly, Figure~\ref{fig:fig9} plots the quantum state fidelity $\mathcal{F}$ as a function of $D/r_0$. The scalar fidelity $\mathcal{F}_{\text{scalar}}$ decays exponentially toward zero in accordance with Equation~(\ref{eq:survival_prob}), while the vector vortex fidelity $\mathcal{F}_{\text{VVB}}$ exhibits only marginal degradation, remaining above $90\%$ across the entire operating range.

\section{Discussion}
The physical mechanism enabling the turbulence resilience of vector vortex beams stems from the isotropic nature of atmospheric refractive index fluctuations. Under standard terrestrial conditions, optical anisotropy induced by atmospheric stress, temperature variations, and magnetic fields is negligible ($\Delta n = |n_x - n_y| < 10^{-9}$) \cite{AndrewsPhillips2005, Kolmogorov1941}. As demonstrated mathematically in Equation~(\ref{eq:phase_diff}), when orthogonal polarization states traverse identical refractive index variations, they acquire an identical scalar phase delay $\Phi_{\text{turb}}(x,y)$. Because the quantum information is encoded in the relative degree of freedom between the two components, the common-mode perturbation vanishes upon differential or Stokes-basis measurement \cite{Gori2001, DAmbrosio2012}.

A crucial finding of this study is that the VVB channel does not exhibit zero error, but rather stabilizes at a residual QBER baseline of approximately $4.8\%$. This error floor is completely independent of the turbulence strength $D/r_0$ and originates from non-turbulent hardware imperfections:
\begin{itemize}
    \item \textbf{SLM pixelation and phase quantization:} Discrete 8-bit quantization and finite pixel pitch ($15\ \mu\text{m}$) introduce higher-order diffraction artifacts during complex holographic beam synthesis via Equation~(\ref{eq:arrizon}) \cite{Arrizon2007, RosalesGuzman2017}.
    \item \textbf{Crystal-induced astigmatism:} Residual birefringence and spatial walk-off in the nonlinear crystal introduce small spatial asymmetries between the paired photons \cite{Nikogosyan1991}.
    \item \textbf{Finite receiver aperture and modal overlap:} Incomplete spatial integration over the detector pupil introduces minor state overlap errors during projective measurements.
\end{itemize}
These limitations are purely technical and can be mitigated by utilizing continuous phase devices, higher-resolution spatial light modulators, or dedicated mode converters and geometric phase metasurfaces \cite{Milione2015, Zhan2009}.

From an operational perspective, this passive resilience provides substantial advantages over active compensation systems. Standard adaptive optics architectures require high-speed wavefront sensors, deformable mirrors, and real-time control loops running at kilohertz frequencies. These systems add weight, optical insertion losses, and latency that degrade the secret-key generation rate. In contrast, the VVB approach provides continuous, latency-free immunity directly embedded in the topological symmetry of the quantum state, making it exceptionally attractive for mobile transceivers, airborne systems, and satellite-to-ground quantum communication downlinks \cite{Willner2015, Yin2017}.

\section{Conclusion}
In conclusion, we have investigated the propagation dynamics and error scaling of hybrid polarization-OAM vector vortex beams in turbulent free-space quantum channels. By exploiting the scalar nature of atmospheric refractive index fluctuations, the non-separable VVB architecture enables complete common-mode cancellation of turbulent phase distortions. Consequently, the local polarization azimuth and normalized Stokes parameters maintain their topological structure even when the underlying scalar wavefront is completely scrambled.

Quantitative numerical simulations across turbulence strengths up to $D/r_0 = 3.0$ demonstrate that while scalar OAM modes suffer catastrophic crosstalk that drives the QBER to $50\%$, the VVB channel suppresses error rates below $4.8\%$, easily satisfying the $11.0\%$ threshold for unconditional QKD security. Yielding an error suppression factor of $11.6$ at $D/r_0 = 1.2$ without adaptive optics, this passive, hardware-free encoding framework provides a robust foundation for long-distance, high-capacity free-space quantum key distribution.

\section*{Acknowledgments}
The authors gratefully acknowledge the Department of Physics at Shahid Beheshti University, Tehran, Iran, for financial and technical support. B.~Talari expresses sincere gratitude to Prof.~Rouhollah Karimzadeh for invaluable academic guidance, scientific supervision, and fruitful discussions throughout this research.

\end{document}